\documentclass{article}
\begin{document}

\begin{titlepage}
\begin{flushright} December 1999\\
 ULB-TH/99-35\\
hep-th/99??
\end{flushright}

\begin{centering}

\vspace{0.5cm}

\huge{A note on the BRST cohomology of the extended antifield formalism} \\

\vspace{.5cm}

\large{Proceedings of the spring school 
{\em Q.F.T., Supersymmetry and Superstrings}, in C\v{a}lim\v{a}ne\c{s}ti,
Romania, April 1998.\\
To appear in {\em Annals of the University of Craiova,
Physics Series}, vol. 9 (1999).} \\

\vspace{.5cm}

\large{Glenn Barnich$^*$}\\

Physique Th\'eorique et Math\'ematique,\\ Universit\'e Libre de
Bruxelles,\\
Campus Plaine C.P. 231, B--1050 Bruxelles, Belgium\\

\end{centering}

\vspace{2.5cm}

\begin{abstract}
The relevance of the BRST cohomology of the extended antifield formalism is 
briefly discussed along with standard homological tools needed for
its computation.
\end{abstract}

\vspace{.5cm}

\vfill

\footnotesize{$^*$Collaborateur scientifique du Fonds National
Belge de la Recherche Scientifique.}

\end{titlepage}

\addtolength{\topmargin}{-2cm}
\addtolength{\textheight}{3.5cm}
\addtolength{\oddsidemargin}{-1cm}
\addtolength{\textwidth}{1.5cm}
\addtolength{\footskip}{0.7cm}
\sloppy

\def\theequation{\thesection.\arabic{equation}}
\newcommand{\mysection}[1]{\section{#1}\setcounter{equation}{0}}
\def\be{\begin{eqnarray}} 
\def\ee{\end{eqnarray}}
\def\beann{\begin{eqnarray*}} 
\def\eeann{\end{eqnarray*}}
\def\beq{\begin{equation}} 
\def\eeq{\end{equation}}
\def\ba{\begin{array}} 
\def\ea{\end{array}}
\def\ben{\begin{enumerate}} 
\def\een{\end{enumerate}}
\def\bea{\begin{eqnarray}} 
\def\eea{\end{eqnarray}}
\def\beann{\begin{eqnarray*}} 
\def\eeann{\end{eqnarray*}}
\def\beq{\begin{equation}} 
\def\eeq{\end{equation}}
\def\ba{\begin{array}} 
\def\ea{\end{array}}
\def\ben{\begin{enumerate}} 
\def\een{\end{enumerate}}

\def\5{\bar }  
\def\6{\partial } 
\def\7{\hat } 
\def\4{\tilde }

\def\gh{\mbox{gh}} 
\def\agh{\mbox{antigh}}
\def\tot{\mbox{totdeg}}
\def\deg{\mbox{formdeg}}

\def\cA{{\cal A}}
\def\cB{{\cal B}}
\def\cC{{\cal C}}
\def\cD{{\cal D}} 
\def\cE{{\cal E}}
\def\cF{{\cal F}}
\def\cG{{\cal G}}
\def\cH{{\cal H}}
\def\cI{{\cal I}}
\def\cJ{{\cal J}}
\def\cK{{\cal K}}
\def\cL{{\cal L}}
\def\cM{{\cal M}}
\def\cN{{\cal N}}
\def\cO{{\cal O}}
\def\cP{{\cal P}}
\def\cQ{{\cal Q}}
\def\cR{{\cal R}} 
\def\cS{{\cal S}}
\def\cT{{\cal T}} 
\def\cU{{\cal U}} 
\def\cV{{\cal V}}
\def\cW{{\cal W}}
\def\cX{{\cal X}}
\def\cY{{\cal Y}}
\def\cZ{{\cal Z}}

\def\s0#1#2{\mbox{\small{$\frac{#1}{#2}$}}}
\def\chris#1#2#3{{\Gamma_{#1#2}}^{#3}}
\def\f#1#2#3{{f_{#1#2}}^{#3}}
\def\fd#1#2#3{{f_{#1#2#3}}}
\def\A#1#2{{A_{#2}}^{#1}}
\def\viel#1#2{e_{#2}{}^{#1}} 
\def\Viel#1#2{E_{#1}{}^{#2}}
\def\csum#1#2{\sum_{#1}\hspace{-1.#2em}\circ\ \ \ }
\def\G{\Gamma}
\def\D{\Delta}

\newtheorem{theorem}{Theorem}
\def\qed{\hbox{${\vcenter{\vbox{
\hrule height 0.4pt\hbox{\vrule width 0.4pt height 6pt
\kern5pt\vrule width 0.4pt}\hrule height 0.4pt}}}$}}

\newtheorem{lemma}{Lemma}

The lectures given at the school 
{\em Q.F.T., Supersymmetry and Superstrings}, in C\v{a}lim\v{a}ne\c{s}ti,
Romania in  April 1998 were intitled 
{\em Classical and quantum 
aspects of the Batalin-Vilkovisky formalism}. 
They covered the
following material: 

\begin{itemize}

\item Lesson 1: Algebraic structure of gauge symmetries

Finite dimensional toy model. 
Noether identities. Koszul-Tate resolution of on-shell functions. 
Gauge symmetries. Longitudinal differential. BRST
differential. Antibracket. Master equation. Homological perturbation
theory. 

\item Lesson 2: Locality in field theory 

Jet-spaces. Local functionals and local functions. Euler-Lagrange
derivatives. Algebraic Poincar\'e lemma. Local antibracket. Master
equation and BRST differential. 

\item Lesson 3: Consistency conditions on anomalies. Non
renormalization theorems. 

Non minimal sector. Gauge fixing. Generating functionals. Statement of 
quantum action principles. Consistency conditions on
anomalies. Lie-Massey brackets. Higher order cohomological
restrictions. Power counting in the antifield formalism. Beta
functions. First non renormalization theorem. Local version of
Callan-Symanzik equation. Second non renormalization
theorem. Chern-Simons theory. 
\end{itemize} 

Useful review references connected to the material covered here are 
\cite{Zin1,PiRo,Olv,Zin2,Bon,HeTe,GPS,PiSo,Wei1,Wei2} .
Details on specific topics, reflecting the point of view of the
author, can be found in 
\cite{BBH,BHW1,BHW2,Ba1,Ba2,Ba3,Ba4,BBHR} 
and in the literature cited therein. 

The purpose of this note is to discuss briefly the
BRST cohomology of the extended antifield formalism, to give 
some details on exact couples and spectral sequences, and to apply these
concepts in the problem at hand. 
 
\mysection{Introduction}

\subsection{Classical theory}

The Batalin-Vilkovisky formalism \cite{BaVi1,BaVi2} allows one to
formulate the BRST differential \cite{BRS1,BRS2,BRS3,Tyu} controling the 
gauge symmetries under renormalization for generic 
gauge theories. 
The formalism can be extended so as to 
include (non linear) global symmetries (see \cite{BHW1,BHW2} 
and references therein), which is achieved by coupling the 
BRST cohomology classes in negative ghost numbers with constant
ghosts. A further extension including the 
BRST cohomology classes in all the ghost numbers can be constructed
 \cite{Ba3}. Some features of this extension are~:
\begin{itemize}
\item it
allows one to take into account in a systematic way all higher order 
cohomological constraints due to the antibracket maps \cite{Ba1}, 
\item it is the appropriate formalism to prove stability independently of 
power counting restrictions, also called renormalizability in the 
modern sense \cite{GoWe}, in the case of generic gauge theories,
\item an appropriate BRST differential 
can be constructed on the classical and the quantum level, 
even in the case of anomalous theories \cite{Ba4}. 
\end{itemize}

Let us briefly summarize the results of \cite{Ba3}
needed in the following. The extended formalism is obtained by 
first computing a basis for the local BRST cohomology classes.
This basis contains
as a subset those classes that can be obtained from the solution 
$S$ of the master equation by differentiation of $S$ with respect to 
so-called essential coupling constants. 
The additional classes completing the basis  
are then coupled
with the help of new independent coupling constants. 
This action can then be extended by terms of
higher orders in the new couplings in such a way that, if we denote by 
$\xi^A$ all the couplings corresponding to the independent 
BRST cohomology classes, the
corresponding action $S$ satisfies the extended master equation
\be
\frac{1}{2}(S,S)+\Delta_c S=0.\label{1}
\ee
The BRST differential associated to
the solution of the extended master equation is 
\be
\bar s=(S,\cdot)+\Delta^L_c,
\ee
where $\D_c={\partial^R\cdot\over\partial\xi^A}f^A$, while 
$\Delta^L_c=(-)^Af^A{\partial^L\over\partial\xi^A}$, with $f^A$
depending (at least quadratically) on the couplings $\xi$ alone. Both
antiderivations satisfy 
$(\Delta_c)^2=0=(\Delta^L_c)^2$. 
Since there is no dependence on the fields and the
antifields, $\Delta^L_c(A,B)=(\Delta^L_cA,B)+(-)^{A+1}(A,\Delta^L_c
B)$, with the appropriate version holding for the right derivation
$\Delta_c$. 
The local BRST cohomology classes contain
the generators of all the generalized non trivial symmetries of the
theory in negative ghost number, the generalized observables in ghost
number zero, and the anomalies (and anomalies for anomalies) in
positive ghost number. This is the reason why the extended master
equation encodes the invariance of the original action under all
the non trivial gauge and glocal symmetries, their commutator
algebra as well as the antibracket algebra of all the 
local BRST cohomology classes. 

The cohomology of $\bar s$ in the space $F$ of $\xi$ dependent local 
functionals in the fields, the antifields and their derivatives
is isomorphic to
the cohomology of 
\be
s_{\Delta_c}=[\Delta_c,\cdot]
\ee
in the space $G$ of graded right
derivations $\lambda=\frac{\partial^R\cdot}{\partial
  \xi^A}\lambda^A$, with $\lambda^A$ a function of $\xi$ 
alone, $[\cdot,\cdot]$ being the graded commutator for graded right
derivations,
\be
H(\bar s,F)\simeq H(s_{\Delta_c},G)\label{7}.
\ee 
If $\mu$ is a $s_{\Delta_c}$ cocycle, the corresponding $\bar s$ cocycle is
given by $\mu S= \frac{\partial^R S}{\partial
  \xi^B}\mu^B$. 

\subsection{Quantum theory}

In the standard version of the BRST-Zinn-Justin-Batalin-Vilkovisky set-up,
there are two main issues to be considered: stability and anomalies.
  
The problem of stability is 
the question if to every local BRST cohomology class in ghost number
$0$, there corresponds an independent coupling of the action. The
extended formalism solves this problem trivially by construction,
since all cohomology classes have been coupled with independent
couplings. The non trivial part of the formalism is the proof of the
existence of the extended master equation and the associated
differential, which allows to control the symmetries after the
extension. Of course, it will be often convenient in practice 
not to couple all
the local BRST cohomology classes but only a subset needed to
guarantee that the theory is stable. 

In the standard set-up, the question of anomalies is mostly reduced to the
question of the local BRST cohomology in ghost number $1$ and to a
discussion of the coefficients of these cohomology classes. In the
presence of anomalies, there is no differential 
on the quantum level associated to the anomalously broken Zinn-Justin
equation for the effective action. In the extended formalism however, 
because all the local BRST cohomology classes in positive ghost 
numbers have been coupled to the
solution of the master equation,
such a differential exists \cite{Ba3}. Indeed,
the quantum action principle \cite{Low,Lam1,Lam2} applied to 
(\ref{1}) gives
\be
\frac{1}{2}(\G,\G)+\D_c\G=\hbar{\cal A}\circ\G\label{2},
\ee
where $\G$ is the renormalized generating functional for 1PI vertices
associated to the solution $S$ of the extended master equation
and the local functional ${\cal A}$ is an element of 
$F$ in ghost number $1$. Using the result (\ref{7}) 
on the cohomology of $\bar s$, 
one can show \cite{Ba3,Ba4} that, through the addition of local counterterms, 
(\ref{2}) can be written as 
\be
\frac{1}{2}(\G^\infty,\G^\infty)+\D^\infty\G^\infty=0,\label{qsm}
\ee
where ${\G}^\infty$ is associated to the action 
${S}^\infty=S-\Sigma_{k=1}\hbar^k\Sigma_k$ containing local finite
BRST breaking counterterms $\Sigma_k$ and 
$\D^\infty=\D_c+\hbar \D_1 +\hbar^2 \D_2+\dots$ satisfies 
$(\D^\infty)^2=0$.
The associated quantum BRST differential is
\be
s^q=(\G^\infty,\cdot)+(\D^\infty)^L. 
\ee
In the limit $\hbar$ going to zero, we recover both the classical
extended master equation (\ref{1}) and the classical 
differential $\bar s$. 

In the extended antifield formalism, the anomalous Zinn-Justin
equation can thus be written as a functional differential equation for the
renormalized effective action. The derivations $\D_1,\D_2,\dots$ are
guaranteed to exist due to the quantum action principles. They
satisfy a priori cohomological restrictions due to the fact that 
the differential $\D^\infty$ is a formal deformation with deformation
parameter $\hbar$ of the differential $\D_c$. 

For instance, the derivation $\D_1$ is a cocycle of $s_{\Delta_c}$ in ghost
number $1$, because $[\Delta_c,\Delta_1]=0$. This cocyle can 
be assumed to be non trivial, because otherwise, $\Delta_1$ could have 
been absorbed by the counterterm $\Sigma_1$. Hence, non trivial anomalies, 
which correspond in this formalism to non trivial 
deformations of $\Delta_c$, are controled by $H^1(s_{\Delta_c},G)$. 

In the standard way, once care has been taken of the trivial anomalies 
through the counterterms $\Sigma_k$, the remaining infinite and finite 
counterterms are required to belong to $H^0(\bar s,F)\simeq
H^0(s_{\Delta_c},G)$ in order to 
preserve (\ref{qsm}) to that order. 

It is thus of interest to compute the cohomology of $s_{\Delta_c}$.

\mysection{BRST cohomology in the extended antifield
  formalism}

Let $\lambda=\frac{\partial^R\cdot}{\partial\xi^A}\lambda^A$ be a
right derivation. We assume that the $\lambda^A$ are formal power
series in $\xi^A$. In the following, we provide this space with an 
obvious filtration. It will however not have finite length, and for 
particular theories, better filtrations have to be found in order 
to do a complete computation. Since the techniques will be similar, 
it is nevertheless useful to show how they work on this example. 

\subsection{Grading and filtration on the space of right derivations}

Let $N_\xi=\frac{\partial^R}{\partial \xi^A}\xi^A$ be 
the operator counting the
number of $\xi$'s. A general right derivation admits the following
decomposition according to the eigenvalues of $N_\xi$:
$\lambda=\lambda_{-1}+\lambda_0+\lambda_{1}+\dots$, where
$[\lambda_p,N_\xi]= p \lambda_p$. 
Hence, $G$ is a graded space, $G=\oplus_{p=-1} G^p$. (It is actually a 
bigraded space, the other grading, for which $s_{\Delta_c}$ is
homogeneous of degree $1$ being the ghost number.)

The graded right commutator 
satisfies $[[\lambda_m,\mu_n],N_\xi]=(m+n)
[\lambda_m,\mu_n]$. The decomposition of $\Delta_c$
starts at eigenvalue $1$: $\D_c=\D_{c1}+\D_{c2}+\dots$~; the
corresponding decomposition of $s_{\Delta_c}$ being $s_{\Delta_c}=
[\D_{c1},\cdot]+[\D_{c1},\cdot]=\dots\equiv s_1+s_2+\dots$. 
 It follows that the cocycle condition
$s_{\Delta_c}\lambda=0$ decomposes as 
\be
s_1\lambda_{-1}&=0,\cr
s_1\lambda_{0}+s_2\lambda_{-1}&=0,\cr
s_1\lambda_{1}+s_2\lambda_{0}+s_3\lambda_{-1}&=0,\cr
&\vdots,
\ee
while the coboundary condition $\lambda=s_{\Delta_c}\mu$ decomposes
as
\be
\lambda_{-1}=0,\cr
\lambda_{0}=s_1\mu_{-1},\cr
\lambda_{1}=s_1\mu_{0}+s_2\mu_{-1},\cr
\lambda_{2}=s_1\mu_{1}+s_2\mu_{0}+s_3\mu_{-1},\cr
\hspace*{-1cm}\vdots,
\ee
In order to construct the spectral sequence associated to this
problem, we follow \cite{BoTu}. 

Consider the spaces $K_{p}$ of derivations having $N_{\xi}$
degree greater than $p$, i.e., $\lambda\in K_p$ if
$\lambda=\lambda_{p}+\lambda_{p+1}+\dots$. The space of all right
derivations is $G=K_{-1}$, $K_{p+1}\subset K_{p}$ and $s_{\Delta_c}
K_p\subset K_p$. The sequence of spaces $K_p$ is a decreasing
filtration of $G$, with $K_p/K_{p+1}\simeq G^p$.  

We have the short exact sequence\footnote{A diagram is said to be exact if the
  image of a map is equal to the kernel of the next map.}: 
\be
0\longrightarrow \oplus_{p=-1} K_{p+1}\stackrel{i}{\longrightarrow}
  \oplus_{p=-1}K_p\stackrel{j}{\longrightarrow} \oplus_{p=-1} K_p/K_{p+1}
\longrightarrow 0,
\ee
where $\oplus_{p=-1} K_p/K_{p+1}\simeq \oplus_{p=-1}G^p$. 
The following diagram is exact at each corner: 
\begin{eqnarray}
\begin{array}{ccc}
H(s_{\Delta_c},\oplus_{p=-1}K_{p+1})\stackrel{i_0}{\longrightarrow}
H(s_{\Delta_c},\oplus_{p=-1}K_p)\\
k_0\nwarrow\ \swarrow j_0\\
E_0,
\end{array}\label{dec}
\end{eqnarray}
where $E_0=\oplus_{p=-1} K_p/K_{p+1}\simeq \oplus_{p=-1} G^p$. 
In this diagram, $H(s_{\Delta_c},K_{p})$ is defined by the cocycle condition 
$s_{\Delta_c}(\lambda_p+\lambda_{p+1}+\dots)=0$, and the coboundary
condition $\lambda_p+\lambda_{p+1}+\dots=s_{\Delta_c}(\mu_p+\mu_{p+1}
+ ...$. The maps $i_0$ and $j_0$ are induced by $i$ and $j$,
$i_0[\lambda_{p+1}+\lambda_{p+2}+\dots]=[\lambda_{p+1}+\lambda_{p+2}+\dots]$ 
and $j_0[\lambda_p+\lambda_{p+1}+\dots]=[j(\lambda_p+\lambda_{p+1}+\dots)]
=[\lambda_p]$. 
They are well defined, because $i_0$ maps cocycles to cocycles and
coboundaries to coboundaries, while
$j(s_{\Delta_c}(\mu_p+\mu_{p+1}\dots))
\in K_{p+1}$. 
The map $k_0$ is defined by $k_0
[\lambda_p]=[s_{\Delta_c}\lambda_p]$. It does not depend on the choice 
of representative for $[\lambda_p]\in K_p/K_{p+1}$, because
$[s_{\Delta_c}(\lambda_{p+1}+\dots)]=0\in H(s_{\Delta_c},K_{p+1})$. 

Let us check explicitly that this diagram is exact:
\begin{itemize}
\item ${\rm ker}\ j_0$ is given by elements $[\lambda_p+\lambda_{p+1}+...]\in 
H(s_{\Delta_c},K_{p})$
such that $s_{\Delta_c}(\lambda_p+\lambda_{p+1}+...)=0$ and
$\lambda_p=0$. This is the same than $i_{0}H(s,K_{p+1})$, which is
given by $[\lambda_{p+1}+\lambda_{p+2}+\dots]$, with
$s_{\Delta_c}(\lambda_{p+1}+\lambda_{p+2}+...)=0$, the equivalence
relation being the equivalence relation in $H(s,K_{p})$ by definition
of $i_0$. 
\item ${\rm ker}\ k_0$ is given by elements $[\lambda_p]$ such that 
$[s_{\Delta_c}\lambda_p]=0\in H(s_{\Delta_c},K_{p+1})$, i.e. such that 
$s_{\Delta_c}\lambda_p=s_{\Delta_c}(\mu_{p+1}+\mu_{p+2}+....)$. By
the identification
$\lambda_{p+1}=-\mu_{p+1},\lambda_{p+2}=-\mu_{p+2},\dots$, this is
indeed the same than $j_0 H(s_{\Delta_c},K_p)$ given by $[\lambda_p]$
with $s_{\Delta_c}(\lambda_p+\lambda_{p+1}+\dots)=0$. 
\item ${\rm ker}\  i_0$ is given by elements 
$[\lambda_{p+1}+\lambda_{p+2}+...]$
  such that $s_{\Delta_c}(\lambda_{p+1}+\lambda_{p+2}+...)=0$ 
and $\lambda_{p+1}+\lambda_{p+2}+...=s_{\Delta_c}
(\mu_{p}+\mu_{p+1}+...)$, while $k_0[\mu_p]$ is given by 
$[\lambda_{p+1}+\lambda_{p+2}+...]$ of the form 
$[s_{\Delta_c}\mu_{p}]$ so that $\lambda_{p+1}+\lambda_{p+2}+...=
s_{\Delta_c}\mu_{p}+s_{\Delta_c}(\mu_{p+1}+\dots)$, which is indeed
the same.
\end{itemize}

\subsection{Exact couples and associated spectral sequence}
To every exact couple $(A_0,B_0)$, i.e., exact diagram of the form 
\begin{eqnarray}
\begin{array}{ccc}
A_0\stackrel{i_0}{\longrightarrow}
A_0\\
k_0\nwarrow\ \swarrow j_0\\
B_0,
\end{array}\label{dec1}
\end{eqnarray}
one can associated a derived exact couple 
\begin{eqnarray}
\begin{array}{ccc}
A_1\stackrel{i_1}{\longrightarrow}
A_1\\
k_1\nwarrow\ \swarrow j_1\\
B_1.
\end{array}\label{dec2}
\end{eqnarray}
In this diagram, the spaces and maps are defined as follows: 
$A_1=i_0A_0$; $B_1=H(d_0,B_0)$, where $d_0=j_0\circ k_0$ ( $d_0^2=0$ because 
$k_0\circ j_0=0$); for $a_1=i_0a_0$, $i_1 a_1=i_1 (i_0a_0)
=i_0^2a_0$; $j_1 a_1=[j_0a_0]$ (this map is well defined: $j_0a_0$ is a
cocycle, because $k_0\circ j_0=0$, furthermore the map does not depend on
the representative choosen for $a_0$, 
because if $i_0a_0=0$, $a_0=k_0 b_0$ for some $b_0$ and
$j_1 a_1 =[j_0\circ k_0 b_0]=0$); $k_1 [b_0]=k_0b_0$ ($k_0b_0 =i_0a_0$ for
some $b_0$ because $d_0b_0=j_0(k_0b_0)=0$ implies $k_0b_0=i_0a_0$, 
furthermore $k_0 d_0c_0=0$
because $k_0\circ j_0=0$).  

Let us also check explicitly exactness of this diagram:
\begin{itemize}
\item ${\rm ker}\ j_1$ is given by elements $a_1=i_0a_0$ such that
  $[j_0a_0]=0$, i.e., $j_0a_0=j_0k_0 b_0$ and then $a_0-k_0b_0
  =i_0c_0$, 
implying that
  $a_1 =i^2_0c_0$. $i_1
  A_1$ is given by elements $a_1=i_1 c_1 =i^2_0c_0$. 
  It follows that ${\rm ker}\ j_1\subset i_1
  A_1$, while the inverse inclusion follows from $j_0\circ i_0=0$.  
\item ${\rm ker}\ k_1$ is given by elements $[b_0]$ such that 
$k_0b_0=i_0a_0=0$, i.e., such that $b_0=j_0c_0$, for some $c_0$, 
while ${\rm im}\ j_1$ is given by elements $[b_0]$ such that
$[b_0]=[j_0e_0]$, i.e $b_0=j_0(e_0 +k_0f_0)$. It follows that ${\rm ker}\
k_1={\rm im}\ j_1$.
\item ${\rm ker}\  i_1$ is given by elements $a_1=i_0a_0$ such
  that $i_0(i_0a_0)=0$, i.e., $i_0a_0=k_0b_0$ 
(which implies in particular $d_0b_0=0$).  
${\rm im}\ k_1$ is given by elements $a_1 =i_0a_0=k_0b_0$ for some
$b_0$ with $d_0b_0=0$, so both spaces are indeed the same. 
\end{itemize}

Clearly, this construction can be iterated by taking as the starting
exact couple the derived couple. We thus get 
a sequence of exact couples 
\begin{eqnarray}
\begin{array}{ccc}
A_r\stackrel{i_r}{\longrightarrow}
A_r\\
k_r\nwarrow\ \swarrow j_r\\
B_r.
\end{array}\label{decr}
\end{eqnarray}
and the associate spectral sequence
$(B_r,d_r)$, for $r=0,1,\dots$, i.e., spaces $B_r$ and differentials
$d_r$ satisfying $B_{r+1}=H(d_r,B_r)$.

\subsection{Spectral sequence associated to the BRST cohomology of 
  the extended antifield formalism}
Let us now apply the general theory to the case of the exact couple
(\ref{dec}) and give explicitly the differentials $d_r$ and the spaces 
$B_r$ (called $E_r$) in this case for $r=0,1,2,3$.  

We have $E_0=\oplus_{p=-1} K_p/K_{p+1}\simeq \oplus_{p=-1} G^p$. The
differential $d_0$ is defined by $d_0[\lambda_p]_0=j_0[s_{\Delta_c}
\lambda_p]$, where 
$[s_{\Delta_c}\lambda_p]\in H(s_{\Delta_c},K_{p+1})$. It follows that 
$d_0[\lambda_p]_0=[s_1\lambda_p]$.
This means that $E^p_1$ is defined by elements $[[\lambda_p]_0]_1$
with the cocycle condition 
\be
s_1 \lambda_p=0
\ee 
and the coboundary condition
\be
\lambda_p=s_1\mu_{p-1}. 
\ee
Because 
$s_1=\frac{\partial^R}{\partial \xi^A}f^{A}_{BC}\xi^B\xi^C$, and  
$s_1^2=0$ implies that the $f^{A}_{BC}$ are the structure constants of a 
graded Lie algebra, this
group is just a graded version of standard Lie algebra 
(Chevalley-Eilenberg) 
cohomology with representation space the adjoint representation. 

Take now $[[\lambda_p]_0]_1\in E^p_1$. The differential 
$d_1[[\lambda_p]_0]_1=j_1k_1[[\lambda_p]_0]_1=j_1k_0[\lambda_p]_0=j_1
[s_{\Delta_c}\lambda_p]=[j_0i_0^{-1}[s_{\Delta_c}\lambda_p]]_1$. This
means that $[s_{\Delta_c}\lambda_p]$ has to be considered as an
element of $H(s_{\Delta_c},K_{p+2})$ so that 
$d_1[[\lambda_p]_0]_1=[[s_2\lambda_p]_0]_1$.
Hence $E^p_2$ is defined by elements $[[[\lambda_p]_0]_1]_2$ with 
the cocyle condition
\be
s_2\lambda_p+s_1\lambda_{p+1}=0,\\
s_1\lambda_p=0,
\ee
and the coboundary condition
\be
\lambda_p=s_2\mu_{p-2}+s_1\mu_{p-1},\\
0=s_1\mu_{p-2}.
\ee
We thus find that $E_2^p=H^p(s_2,H(s_1))$. 

The differential $d_2$ in $E^p_2$ is defined by 
$d_2[[[\lambda_p]_0]_1]_2=j_2k_2[[[\lambda_p]_0]_1]_2=
j_2k_1[[\lambda_p]_0]_1=j_2k_0[\lambda_p]_0
=[j_1i_1^{-1}k_0[\lambda_p]_0]_2
=[[j_0i_0^{-1}i_1^{-1}k_0[\lambda_p]_0]_1]_2$. 
In order to make sure that $k_0[\lambda_p]_0$ belongs to
$i_1i_0H(s_{\Delta_c},K_{p+1})$ we use 
$\lambda_p+\lambda_{p+1}$ as a representative for $[\lambda_p]_0$. It
follows that $d_2[[[\lambda_p]_0]_1]_2=[[[s_3\lambda_p
+s_2\lambda_{p+1}]_0]_1]_2$. The cocycle condition for an element 
$[[[[\lambda_p]_0]_1]_2]_3\in E^p_3$ is then given by 
\be
s_3\lambda_p+s_2\lambda_{p+1}=s_2\mu_{p+1}+s_1\mu_{p+2},\\
s_2\lambda_p+s_1\lambda_{p+1}=0,\\
s_1\lambda_{p}=0,
\ee
with $s_1\mu_{p+2}=0$. The redefinition
$\lambda_{p+1}\rightarrow\lambda_{p+1}-\mu_{p+1}$ and
$\lambda_{p+2}=-\mu_{p+2}$, then gives as cocycle condition
\be
s_3\lambda_p+s_2\lambda_{p+1}+s_1\lambda_{p+2}=0,\\
s_2\lambda_p+s_1\lambda_{p+1}=0,\\
s_1\lambda_{p}=0.
\ee
The coboundary condition is
$[[[\lambda_p]_0]_1]_2=d_3[[[\mu_{p-3}+\mu_{p-2}]_0]_1]_2$, where
$s_1\mu_{p-3}=0,s_2\mu_{p-3}+s_2\mu_{p-2}=0$, hence
$[[\lambda_p]_0]_1=[[s_3\mu_{p-3}+s_2\mu_{p-2}]_0]_1
+d_2[[\sigma_{p-2}]_0]_1$, 
with $s_1\sigma_{p-2}=0$ which gives 
\be
\lambda_p=s_3\mu_{p-3}+s_2\mu_{p-2}+s_2\sigma_{p-2}+s_1\rho_{p-1},\\
0=s_2\mu_{p-3}+s_1\mu_{p-2},\\
0=s_1 \mu_{p-3},\ 0=s_1\sigma_{p-2}.
\ee
The redefinition $\mu_{p-2}\rightarrow\mu_{p-2}+\sigma_{p-2}$ and 
$\rho_{p-1}=\mu_{p-1}$, then gives the coboundary condition
\be
\lambda_p=s_3\mu_{p-3}+s_2\mu_{p-2}+s_1\mu_{p-1},\\
0=s_2\mu_{p-3}+s_1\mu_{p-2},\\
0=s_1 \mu_{p-3}.
\ee
This construction can be continued in the same way for higher $r$'s.

The original problem was the computation of 
$H(s_{\Delta_c},G)=H(s_{\Delta_c},K_{-1})$.
From exactness of the couples (\ref{decr}), it follows that 
\be
H(s_{\Delta_c},K_{-1})\simeq j_0 E^{-1}_0
\oplus {\rm ker}\ j_0\nonumber\\
\simeq
{\rm ker}\ k_0(\subset E^{-1}_0)\oplus i_0 
H(s_{\Delta_c},K_{0})\nonumber\\
\simeq
{\rm ker}\
k_0(\subset E^{-1}_0)\oplus {\rm ker}\
k_1(\subset E^{0}_1)\oplus i_1 i_0H(s_{\Delta_c},K_{1})\\
\vdots \\
\simeq \oplus_{r=0}^R{\rm ker}\
k_r(\subset E^{r-1}_r)\oplus i_R\dots i_0 H(s_{\Delta_c},K_{R}).
\ee
Furthermore, $E_0\simeq E_1\oplus F_0\oplus d_0 F_0$ 
and $E^{-1}_0\simeq
E^{-1}_1 \oplus F_0^{-1}$. $F_0$ does not belong to ${\rm ker}\ k_0$
because $d_0 F_0\neq 0$. Thus
${\rm ker}\ k_0(\subset E^{-1}_0))\simeq {\rm ker}\ k_1(\subset
E^{-1}_1)$. Similarily, $E_1\simeq E_2\oplus F_1\oplus d_1 F_1$ and 
$(d_1 F_1)^{-1}=(d_1 F_1)^{0}=0$. Again, $d_1[F_1]_1\neq 0$ implies
that $F_1$ does not belong to  ${\rm ker}\ k_1$. This means that 
${\rm ker}\ k_1(\subset
E^{-1}_1)\simeq {\rm ker}\ k_2(\subset
E^{-1}_2)$ and ${\rm ker}\ k_1(\subset
E^{0}_1)\simeq {\rm ker}\ k_2(\subset
E^{0}_2)$.
Going on in the same way, we conclude that 
${\rm ker}\ k_r(\subset E^{r-1}_r))\simeq {\rm ker}\ k_R(\subset
E^{r-1}_R))$.
We thus get 
\be
H(s_{\Delta_c},K_{-1})\simeq \oplus_{r=0}^R {\rm ker}\
k_R(\subset E^{r-1}_R)\oplus i_R\dots i_0 H(s_{\Delta_c},K_{R}).
\ee
 
This construction is most useful 
if it would stop at some point. Indeed, suppose that $K_{R}=0$. 
Because $k_R[\dots[\lambda_p]_0\dots]_R$
belongs to $i_{R}\dots i_0 H(s_{\Delta_c},K_{p+R+1})=0$, it follows
that 
\be
H(s_{\Delta_c},K_{-1})\simeq \oplus_{r=0}^{R-1} E^{r-1}_R.
\ee

\section*{Acknowledgments}

The author wants to thank the F.N.R.S. (Belgium) for travel support, 
the organizers of the school, 
Radu Constantinescu and Florea Uliu of the Physics Department, 
University of Craiova and Mihail Sandu from the Economic
High School, Calimanesti, for the opportunity to lecture. 
He also thanks the official sponsors of the school, among them 
the director and staff of the Economic
High School, Calimanesti, for 
the warm welcome in Romania. 
He acknowlegdes useful discussions with 
F.~Brandt on the material presented in this note. 
This work has been partly supported by the ``Actions de
Recherche Concert{\'e}es" of the ``Direction de la Recherche
Scientifique - Communaut{\'e} Fran{\c c}aise de Belgique", by
IISN - Belgium (convention 4.4505.86). 

\vfill
\pagebreak

\end{document}